# Multi-component lattice-Boltzmann model with interparticle interaction

Xiaowen Shan[1,2] and Gary Doolen[2]

A previously proposed [X. Shan and H. Chen, Phys. Rev. E **47**, 1815, (1993)] lattice Boltzmann model for simulating fluids with multiple components and interparticle forces is described in detail. Macroscopic equations governing the motion of each component are derived by using Chapman-Enskog method. The mutual diffusivity in a binary mixture is calculated analytically and confirmed by numerical simulation. The diffusivity is generally a function of the concentrations of the two components but independent of the fluid velocity so that the diffusion is Galilean invariant. The analytically calculated shear kinematic viscosity of this model is also confirmed numerically.

**KEY WORDS:** lattice-Boltzmann; multi-phase flow; diffusion.

## 1. Introduction

Recently the lattice-Boltzmann Equation (LBE) method has emerged as a new promising method of computational fluid mechanics (CFD). This method was developed from a discretized fluid model known as the Lattice Gas Automaton (LGA).[1,2] In the LGA model, fluid is modeled microscopically as a collection of particles moving on a regular lattice along the links. The particles collide with each other on lattice sites according to some carefully designed collision rules which conserve the number of particles and momentum. The coarse-grained fluid variables, such as density and fluid velocity, can be shown to obey a set of macroscopic equations which are very similar to the Navier-Stokes

---

[1] US Air Force, Phillips Laboratory, Hanscom Field, Massachusetts, 01731
[2] Center for Nonlinear Studies, Los Alamos National Laboratory, Los Alamos, NM 87545

equations. Since only a small number of bits are required to characterize the states of each lattice site, and the collision operation is local in most cases, the LGA can be implemented very efficiently for a large number of lattice sites. Fluid motion can then be simulated without integrating partial differential equations.

Although the LGA method has a few highly desirable advantages over the conventional methods, it also suffers from intrinsic drawbacks. For instance, the large statistical fluctuation limits the practical usage of the LGA method. To suppress the statistical noise, McNamara and Zanetti[3] suggested modeling the LGA with a lattice Boltzmann equation. Mean population, instead of the discrete particles, are used to simulate fluid flows. A similar approach was proposed by Higuera et al[4] in which a linearized collision term is used in the lattice Boltzmann equation so that the algorithm is computationally more efficient and can be easily generalized to three dimensions. More recently, it was pointed out[5,6] that in addition to the suppression of statistical noise, the unphysical artifacts in the original LGA, known as the lack of Galilean invariance and the velocity-dependent pressure term, can also be eliminated when a single relaxation time collision term (also known as the BGK collision term after Bhatnagar, Gross and Krook[7]) with a proper choice of the equilibrium distribution function is adopted in the lattice Boltzmann equation. Comparisons with conventional CFD methods demonstrated that such a LBE model can give quite satisfactory results in simulating both hydrodynamics and magnetohydrodynamics problems.[8,9]

An interesting and important application of the LGA/LBE methods is the simulation of fluid flows with interfaces between multiple phases. There are numerous complex flow systems in both natural and industrial processes that involve convection-coupled mass transfer near fluid interfaces. The density gradients in such problems are often so large that the conventional linear diffusion equation cease to be an acceptable approximation. Such problems have posed considerable difficulties to the conventional CFD methods, especially when the interfaces can undergo topological changes. Since the formation of fluid interfaces is microscopically due to the long-range interaction between the molecules of the fluid,[10] the separation and reconnection of the interfaces require additional terms to be inserted in the Navier-Stokes equations. While the method of Molecular Dynamics can treat interfacial problems by taking into account the details of the inter-molecular interaction, it becomes computationally too expansive when the large scale flow structures have to be modeled simultaneously.

Because the LGA and LBE methods were developed based on microscopic description of fluids, interactions between fluid elements can be naturally included. Several LGA/LBE models for multi-phase flows have been developed since the first introduction of the LGA. Rothman and Keller[11] developed the first LGA model for simulation of two immiscible fluids. A Boltzmann equation version was formulated later.[12] The same idea was used to achieve a low diffusivity in the simulation of partially miscible fluids[13] with LBE method. In these models, a repulsive force between the two fluid components is most intense in the interfacial zone where the "color gradient" is large. Since the "interfacial zone" is defined by an arbitrary small parameter, for the LBE model, a finite-amplitude perturbation is required for a phase separation to occur in a initially homogeneous system. The early stages of phase separation can be expected to have some arbitrariness. In addition, these models have other problems such as the anisotropy of the surface tension[14,12] and difficulties in dealing with components with different densities. Appert et al[15] suggested another LGA model to simulate a liquid-gas type of phase transition. Attractive or repulsive forces exist between particles that are several lattice units apart. As the range of the force increases, the system separates into two phases. However, the low efficiency and all the problems that plague the LGA methods have prevented this model from being used to solve practical problems.

In a previous paper,[16] we proposed a scheme to incorporate the interparticle forces in LBE models with multiple components. Interaction potentials of different nature are defined between particles of the same or different components. Each component has its own molecular mass and relaxation time. The LBE model was shown to be able to simulate a fluid with an arbitrary non-ideal gas equation of state. When the equation of state is properly chosen, phase separation occurs both in single- and multiple-component systems. In most cases, the interaction can be restricted to involve only nearest neighbors so that the model is computationally efficient. Isotropy of the surface tension and the density profile across the liquid-gas interface have been shown both analytically and numerically in a single component fluid containing two phases.[17] In order for this model to be used in quantitative simulations of multiphase flow problems, the densities of the two phases have also been obtained analytically as functions of a temperature-like parameter.

In a system with more than one component, there could be a very complicated dependence on the details of the interaction potential. Several phases with different densities and concentrations and multiple diffusion processes exist in such a system. In general, the gradients of the concentrations can be very large and the diffusivity depends on the concentration of each component. Before the LBE model can be used to simulate multiphase flow with mass transfer between phases, the diffusion process has to be well understood and the diffusivity calculated. In this article, we give the details of the multiple component LBE model with interparticle forces. In Section 2, we derive the macroscopic fluid equations governing the motion of each component using the Chapman-Enskog expansion method. We obtain the mutual diffusivity in a binary mixture as a function of the concentrations of the components. It does not depend on the fluid velocity. In Section 3, the diffusivity in a numerical experiment on a two-dimensional hexagonal lattice was measured and found to be in excellent agreement with the analytical values. Problems involving convection-diffusion processes near a two-liquid interface and evaporation in porous media have been investigated using the present LBE model. These simulation results will be discussed in future publications.

**2. Multiple component Lattice Boltzmann model**

We now review the multiple-component LBE model[16] with interparticle forces. Consider the motion of particles of $S$ different components on a regular lattice in $D$-dimensional space. The particles of the $\sigma$th component have the molecular mass of $m_\sigma$, $\sigma = 1, \ldots, S$. The population of the particles of the $\sigma$th component having the velocity $\mathbf{e}_a$ at lattice site $\mathbf{x}$ and time $t$ is denoted by $n_a^\sigma(\mathbf{x}, t)$; where, $\{\mathbf{e}_a; a = 1, \ldots, b\}$ is the set of vectors of length $c$ pointing from $\mathbf{x}$ to its $b$ neighboring sites. The evolution of $n_a^\sigma(\mathbf{x}, t)$ is described by the following lattice Boltzmann equations:

$$n_a^\sigma(\mathbf{x} + \mathbf{e}_a, t + 1) - n_a^\sigma(\mathbf{x}, t) = -\frac{1}{\tau_\sigma}\left[n_a^\sigma(\mathbf{x}, t) - n_a^{\sigma(\text{eq})}(\mathbf{x}, t)\right]; \quad \sigma = 1, \cdots, S. \tag{1}$$

where $\tau_\sigma$ is the collision interval of the $\sigma$th component. On the right hand side, we adopt the BGK collision term and chose the equilibrium distribution functions to be $n_a^{\sigma(\text{eq})} = N_a(n_\sigma, \mathbf{u}_\sigma^{\text{eq}})$, where

$$N_a(n, \mathbf{u}) = \begin{cases} n\left(\frac{1-d_0}{b} + \frac{D}{c^2 b}\mathbf{e}_a \cdot \mathbf{u} + \frac{D(D+2)}{2c^4 b}\mathbf{e}_a\mathbf{e}_a : \mathbf{u}\mathbf{u} - \frac{D\mathbf{u}^2}{2c^2 b}\right); & a = 1, \cdots, b \\ n\left(d_0 - \frac{\mathbf{u}^2}{c^2}\right); & a = 0 \end{cases}. \tag{2}$$

$d_0 > 0$ is an arbitrary constant. It was shown that this choice of the equilibrium distribution corrects the unphysical artifacts of the LGA and yields the correct Navier-Stokes equations.[5] In the equations above, $n_\sigma = \sum_a n_a^\sigma$ is the number density of the $\sigma$th component. $\mathbf{u}_\sigma^{\text{eq}}$ are parameters which will be calculated from the distribution functions and the long-range interactions.

In the absence of interparticle forces, all the components are ideal gases. We assume that the velocities $\mathbf{u}_\sigma^{\text{eq}}$, $\sigma = 1, \ldots, S$, all equal a common velocity $\mathbf{u}'$. Since in this case the total momentum of particles of all components should be conserved by the collision operator at each lattice site, it follows directly from Eq. (1) that

$$\mathbf{u}' = \sum_\sigma \frac{\rho_\sigma \mathbf{u}_\sigma}{\tau_\sigma} \bigg/ \sum_\sigma \frac{\rho_\sigma}{\tau_\sigma}, \qquad (3)$$

where $\rho_\sigma = m_\sigma n_\sigma$ is the $\sigma$th component mass density, and $\rho_\sigma \mathbf{u}_\sigma = m_\sigma \sum_a n_a^\sigma \mathbf{e}_a$ is the $\sigma$th component momentum.

We let the strength of the long range force between particles of component $\sigma$ at site $\mathbf{x}$ and particles of component $\bar{\sigma}$ at site $\mathbf{x}'$ be proportional to the product of their "effective masses," $\psi_\sigma(n_\sigma(\mathbf{x}))\psi_{\bar{\sigma}}(n_{\bar{\sigma}}(\mathbf{x}'))$. The "effective mass" of the $\sigma$th component $\psi_\sigma(n_\sigma(\mathbf{x}))$ is defined as a function of the local density $n_\sigma$. The form of $\psi_\sigma(n)$ can be arbitrarily chosen and will determine the equations of state of the fluid components and composite fluid.[17] Summing over all the components and interacting sites, the total long range force acting on the particles of the $\sigma$th component at site $\mathbf{x}$ is:

$$\mathbf{F}_\sigma(\mathbf{x}) = -\psi_\sigma(\mathbf{x}) \sum_{\mathbf{x}'} \sum_{\bar{\sigma}} G_{\sigma\bar{\sigma}}(\mathbf{x}, \mathbf{x}') \psi_{\bar{\sigma}}(\mathbf{x}')(\mathbf{x}' - \mathbf{x}), \qquad (4)$$

where the Green's function satisfies $G_{\sigma\bar{\sigma}}(\mathbf{x}, \mathbf{x}') = G_{\bar{\sigma}\sigma}(\mathbf{x}', \mathbf{x})$. If only homogeneous isotropic interactions between the nearest neighbors are allowed, $G_{\sigma\bar{\sigma}}(\mathbf{x}, \mathbf{x}')$ reduces to the following symmetric matrix with constant elements,

$$G_{\sigma\bar{\sigma}}(\mathbf{x}, \mathbf{x}') = \begin{cases} 0, & |\mathbf{x} - \mathbf{x}'| > c \\ \mathcal{G}_{\sigma\bar{\sigma}}, & |\mathbf{x} - \mathbf{x}'| \leq c. \end{cases} \qquad (5)$$

Furthermore, if the densities $n_\sigma$ vary at a spatial scale much larger than the lattice spacing, $c$, $\mathbf{F}_\sigma$ can be approximated by:

$$\mathbf{F}_\sigma \simeq -\frac{c^2 b}{D} \psi_\sigma \sum_{\bar{\sigma}} \mathcal{G}_{\sigma\bar{\sigma}} \nabla \psi_{\bar{\sigma}}. \qquad (6)$$

We limit the discussion in this paper to interactions which are homogeneous, isotropic, and between nearest neighbors.

The long-range interparticle force introduced above causes an extra momentum change to the $\sigma$th component in addition to the momentum exchange caused by collisions with other components. To incorporate this momentum change in the dynamics of the distribution functions, one can simply define

$$\rho_\sigma \mathbf{u}_\sigma^{\text{eq}} = \rho_\sigma \mathbf{u}' + \tau_\sigma \mathbf{F}_\sigma. \tag{7}$$

It can be verified that at each time step, the momentum change of the $\sigma$th component due to long range interaction is $\mathbf{F}_\sigma$.

When a long-range force exists, the collision operator does not locally conserve the total momentum of all the components, although it does conserve the number of particles of each component. By summing Eq. (1) over all directions, we obtain a mass-conservation relation for each of the $S$ components. Multiplying Eq. (1) by $\mathbf{e}_a$ and $m_\sigma$, and summing over all the $b$ directions and $S$ components, we can obtain the change of total momentum at each site. These relations are:

$$\sum_a n_a^\sigma(\mathbf{x}+\mathbf{e}_a, t+1) - \sum_a n_a^\sigma(\mathbf{x}, t) = 0; \qquad \sigma = 1, \ldots, S, \tag{8}$$

$$\sum_\sigma m_\sigma \sum_a n_a^\sigma(\mathbf{x}+\mathbf{e}_a, t+1)\mathbf{e}_a - \sum_\sigma m_\sigma \sum_a n_a^\sigma(\mathbf{x}, t)\mathbf{e}_a = \sum_\sigma \mathbf{F}_\sigma. \tag{9}$$

The second equation states that during a collision, the total momentum of the particles at a lattice site is changed by the interactions with the particles on neighboring sites. Nevertheless, the momentum of the whole system can be shown to be conserved. We find the change of total momentum of the whole system by summing over all the lattice sites,

$$\Delta \mathbf{P} = \sum_\mathbf{x} \sum_\sigma \mathbf{F}_\sigma(\mathbf{x}) = -\sum_\mathbf{x} \sum_{\sigma,\bar\sigma} \mathcal{G}_{\sigma\bar\sigma} \sum_a \psi_\sigma(\mathbf{x})\psi_{\bar\sigma}(\mathbf{x}+\mathbf{e}_a)\mathbf{e}_a. \tag{10}$$

Here, $\mathbf{e}_a$ is a dummy variable. If there is no net momentum flux at the boundaries, as in the case with periodic boundaries, $\mathbf{x}$ can be viewed as a dummy variable too. After changes of variable $\mathbf{e}_a \to -\mathbf{e}_a$ and then $\mathbf{x}-\mathbf{e}_a \to \mathbf{x}$, we have

$$\Delta \mathbf{P} = \sum_\mathbf{x} \sum_{\sigma,\bar\sigma} \mathcal{G}_{\bar\sigma\sigma} \sum_a \psi_\sigma(\mathbf{x})\psi_{\bar\sigma}(\mathbf{x}+\mathbf{e}_a)\mathbf{e}_a. \tag{11}$$

Since the matrix $\mathcal{G}_{\sigma\bar{\sigma}}$ is symmetric, we have $\Delta \mathbf{P} = -\Delta \mathbf{P}$, and therefore $\Delta \mathbf{P} = 0$.

The question arises as how to calculate the macroscopic "fluid velocity" from the distribution functions since the momentum of the $\sigma$th component are $m_\sigma \sum_a n_a^\sigma \mathbf{e}_a$ before a collision and $(1 - \frac{1}{\tau_\sigma}) m_\sigma \sum_a n_a^\sigma \mathbf{e}_a + \frac{m_\sigma}{\tau_\sigma} \sum_a n_a^{\sigma(\text{eq})} \mathbf{e}_a$ after collision, as obtained by summing the distribution function before and after collision. They differ by a significant amount in region containing large density gradients such as the interface between two phases. Since it is the average of the two values that represents the over-all mass transfer, we define the velocity of the whole fluid, $\mathbf{u}$, to be given by the average of the two momentum values summed over all the components. We can show that this fluid velocity vanishes at equilibrium. Using Eqs. (7) and (3), we find the average to be

$$\rho \mathbf{u} = \sum_\sigma m_\sigma \sum_a n_a^\sigma \mathbf{e}_a + \frac{1}{2} \sum_\sigma \mathbf{F}_\sigma, \tag{12}$$

where $\rho = \sum_\sigma \rho_\sigma$ is the total mass density of the fluid. The second term is generally negligible except in the interfacial zone.

We now follow the Chapman-Enskog method of successive approximation[18,19] to obtain the macroscopic fluid equation of the multiple-component LBE model. The equilibrium distribution functions, $n_a^\sigma$, are expanded as an infinite series, and the time-derivative $\partial/\partial t$ is also divided into parts accordingly:

$$n_a^\sigma = \sum_{r=0}^\infty n_a^{\sigma(r)}; \quad \frac{\partial}{\partial t} = \sum_{r=1}^\infty \frac{\partial_r}{\partial t}. \tag{13}$$

Since now the collision operator depends upon the spatial derivatives of the densities, it is impossible to choose a distribution function which is function of velocity and density only and makes the collision term vanish. We therefore let the leading order distribution functions of all components be the equilibrium distribution about the fluid velocity, $\mathbf{u}$, namely $n_a^{\sigma(0)} = N_a(n_\sigma, \mathbf{u})$, so that a set of macroscopic equations in terms of the correct fluid variables can be obtained as the result of the expansion procedure. The Boltzmann equation will be satisfied at the next order when terms that depend on spatial derivatives are included. Because $\sum_a n_a^{\sigma(0)} = n_\sigma$ and $\sum_\sigma m_\sigma \sum_a n_a^{\sigma(0)} \mathbf{e}_a = \rho \mathbf{u}$, the next order terms will satisfy the following equations:

$$\sum_a n_a^{\sigma(1)} = 0; \quad \sum_\sigma m_\sigma \sum_a n_a^{\sigma(1)} \mathbf{e}_a = -\frac{1}{2} \sum_\sigma \mathbf{F}_\sigma. \tag{14}$$

Taylor-expanding the left-hand side of Eq. (1) to second order and using the expansions (13), we have

$$n_a^\sigma(\mathbf{x}+\mathbf{e}_a, t+1) - n_a^\sigma(\mathbf{x},t) \simeq \frac{\partial_1 n_a^{\sigma(0)}}{\partial t} + \mathbf{e}_a \cdot \nabla n_a^{\sigma(0)} + \frac{\partial_2 n_a^{\sigma(0)}}{\partial t} + \frac{\partial_1 n_a^{\sigma(1)}}{\partial t}$$
$$+ \mathbf{e}_a \cdot \nabla n_a^{\sigma(1)} + \frac{1}{2}\frac{\partial_1}{\partial t}\left(\frac{\partial_1 n_a^{\sigma(0)}}{\partial t} + \mathbf{e}_a \cdot \nabla n_a^{\sigma(0)}\right) + \frac{1}{2}\mathbf{e}_a \cdot \nabla\left(\frac{\partial_1 n_a^{\sigma(0)}}{\partial t} + \mathbf{e}_a \cdot \nabla n_a^{\sigma(0)}\right) \quad (15)$$

On substituting Eq. (15) into the conservation relations Eqs. (8)–(9) and collecting terms of leading order, the following equations are obtained:

$$\frac{\partial_1 n_\sigma}{\partial t} + \nabla \cdot (n_\sigma \mathbf{u}) = 0, \qquad \sigma = 1, \cdots, S \quad (16)$$

$$\frac{\partial_1 \rho \mathbf{u}}{\partial t} + \nabla \cdot \left[\frac{c^2(1-d_0)}{D}\rho + \rho \mathbf{u}\mathbf{u}\right] = \sum_\sigma \mathbf{F}_\sigma. \quad (17)$$

Using Eq. (16), and noting that $\sum_\sigma \mathbf{F}_\sigma \simeq -\frac{c^2 b}{2D}\nabla \sum_{\sigma\bar\sigma}\mathcal{G}_{\sigma\bar\sigma}\psi_\sigma\psi_{\bar\sigma}$, Eq. (17) can be written in the following form

$$\frac{D_1 \mathbf{u}}{Dt} = -\frac{1}{\rho}\nabla p, \quad (18)$$

where $D_1/Dt = \partial_1/\partial t + \mathbf{u} \cdot \nabla$, and $p$ is the pressure given as function of the densities of all the components by the following equation of state

$$p = \frac{c^2}{D}\left[(1-d_0)\rho + \frac{b}{2}\sum_{\sigma\bar\sigma}\mathcal{G}_{\sigma\bar\sigma}\psi_\sigma\psi_{\bar\sigma}\right]. \quad (19)$$

The second second term in Eq. (19), which causes the equation of state of the fluid to be one of a non-ideal gas, depends explicitly on the interparticle force.

The terms of the next order in the expansion of the mass-conservation relation yield the following equation for each component:

$$\frac{\partial_2 \rho_\sigma}{\partial t} + m_\sigma \nabla \cdot \sum_a n_a^{\sigma(1)}\mathbf{e}_a + \frac{m_\sigma}{2}\nabla \cdot \left(\frac{\partial_1}{\partial t}\sum_a n_a^{\sigma(0)}\mathbf{e}_a + \nabla \cdot \sum_a n_a^{\sigma(0)}\mathbf{e}_a\mathbf{e}_a\right) = 0. \quad (20)$$

Using Eq. (16), we can evaluate the third term.

$$m_\sigma \left( \frac{\partial_1}{\partial t} \sum_a n_a^{\sigma(0)} \mathbf{e}_a + \nabla \cdot \sum_a n_a^{\sigma(0)} \mathbf{e}_a \mathbf{e}_a \right) = -\frac{\rho_\sigma}{\rho} \nabla p + \frac{c^2(1-d_0)}{D} \nabla \rho_\sigma. \quad (21)$$

To calculate $\sum_a m_\sigma n_a^{\sigma(1)} \mathbf{e}_a$ in the second term of (21), we substitute the expansions of Eq (13) up to second order into the kinetic equation (1), obtaining the following equations:

$$\frac{\partial_1}{\partial t} \sum_a m_\sigma n_a^{\sigma(0)} \mathbf{e}_a + \nabla \cdot \sum_a m_\sigma n_a^{\sigma(0)} \mathbf{e}_a \mathbf{e}_a = -\frac{1}{\tau_\sigma} \left[ \rho_\sigma (\mathbf{u} - \mathbf{u}') + \sum_a m_\sigma n_a^{\sigma(1)} \mathbf{e}_a \right] + \mathbf{F}_\sigma, \quad (22)$$

which when multiplied by $\tau_\sigma$ and summed over all the components, become

$$\rho(\mathbf{u} - \mathbf{u}') = \sum_\sigma \tau_\sigma \mathbf{F}_\sigma + \frac{1}{2} \sum_\sigma \mathbf{F}_\sigma + \sum_\sigma \frac{\tau_\sigma \rho_\sigma}{\rho} \nabla p - \frac{c^2(1-d_0)}{D} \sum_\sigma \tau_\sigma \nabla \rho_\sigma. \quad (23)$$

In the derivation above, Eqs. (14) and (21) are used. The expression $\sum_a m_\sigma n_a^{\sigma(1)} \mathbf{e}_a$ can then be solved from Eq. (22) in terms of the macroscopic fluid variables. Combining Eq. (20) with the first order equation (16), we obtain the following equations at second order

$$\frac{\partial \rho_\sigma}{\partial t} + \nabla \cdot (\rho_\sigma \mathbf{u}) = -\tau_\sigma \nabla \cdot \mathbf{F}_\sigma + \left( \tau_\sigma - \frac{1}{2} \right) \nabla \cdot \left[ \frac{c^2(1-d_0)}{D} \nabla \rho_\sigma - \frac{\rho_\sigma}{\rho} \nabla p \right]$$
$$+ \nabla \cdot \frac{\rho_\sigma}{\rho} \left[ \sum_\sigma \tau_\sigma \mathbf{F}_\sigma + \frac{\nabla p}{\rho} \sum_\sigma \tau_\sigma \rho_\sigma + \frac{1}{2} \sum_\sigma \mathbf{F}_\sigma - \frac{c^2(1-d_0)}{D} \sum_\sigma \tau_\sigma \nabla \rho_\sigma \right]. \quad (24)$$

When summed over all the components, the equations above become the continuity equation of the whole fluid at second order:

$$\frac{\partial \rho}{\partial t} + \nabla \cdot (\rho \mathbf{u}) = 0. \quad (25)$$

Mutual diffusivity in a two-component system can be calculated using Eq. (24). For notational convenience, we rescale the interaction constants $\mathcal{G}_{\sigma\bar{\sigma}}$ as $G_{\sigma\bar{\sigma}} = b\mathcal{G}_{\sigma\bar{\sigma}}/(1-d_0)$. Using Eq. (19), Eq. (24) can be written as

$$\frac{\partial \rho_1}{\partial t} + \nabla \cdot (\rho_1 \mathbf{u}) = \frac{c^2(1-d_0)}{D} \nabla \cdot (A \nabla \rho_1 - B \nabla \rho_2) \quad (26)$$

$$\frac{\partial \rho_2}{\partial t} + \nabla \cdot (\rho_2 \mathbf{u}) = \frac{c^2(1-d_0)}{D} \nabla \cdot (B \nabla \rho_2 - A \nabla \rho_1). \quad (27)$$

Here $A$ and $B$ are functions of the densities of the two components

$$A = \frac{\rho_2}{\rho}\left(\frac{\rho_1\tau_2 + \rho_2\tau_1}{\rho} - \frac{1}{2}\right) + \psi_1'\frac{\rho_1\tau_2 + \rho_2\tau_1}{\rho}\frac{G_{11}\psi_1\rho_2 - G_{12}\psi_2\rho_1}{\rho} \quad (28)$$

$$B = \frac{\rho_1}{\rho}\left(\frac{\rho_1\tau_2 + \rho_2\tau_1}{\rho} - \frac{1}{2}\right) + \psi_2'\frac{\rho_1\tau_2 + \rho_2\tau_1}{\rho}\frac{G_{22}\psi_2\rho_1 - G_{12}\psi_1\rho_2}{\rho}, \quad (29)$$

where $\psi_\sigma = \psi_\sigma(\rho_\sigma^0)$, and $\psi_\sigma' = d\psi_\sigma/d\rho_\sigma|_{\rho_\sigma = \rho_\sigma^0}$.

We consider the evolution of small perturbations of the densities. Let $\rho_\sigma = \rho_\sigma^0 + \rho_\sigma^1 + \cdots$, where $\rho_\sigma^0$ are uniform, time-independent equilibrium densities of the two components, and $\rho_\sigma^1 \ll \rho_\sigma^0$ are small perturbations. Define the concentration of the $\sigma$th component as $c_\sigma = \rho_\sigma^0/\rho^0$, where $\rho^0 = \rho_1^0 + \rho_2^0$. Obviously $c_1 + c_2 = 1$. We linearize Eqs. (26)–(27) about the equilibrium densities to obtain

$$\frac{D\rho_1^1}{Dt} + \rho_1^0 \nabla \cdot \mathbf{u} = \frac{c^2(1-d_0)}{D}\nabla \cdot \left(A^0 \nabla \rho_1^1 - B^0 \nabla \rho_2^1\right) \quad (30)$$

$$\frac{D\rho_2^1}{Dt} + \rho_2^0 \nabla \cdot \mathbf{u} = \frac{c^2(1-d_0)}{D}\nabla \cdot \left(B^0 \nabla \rho_2^1 - A^0 \nabla \rho_1^1\right) \quad (31)$$

where $D/Dt = \partial/\partial t + \mathbf{u} \cdot \nabla$, and

$$A^0 = c_2(c_1\tau_2 + c_2\tau_1 - \frac{1}{2}) + \psi_1'(c_1\tau_2 + c_2\tau_1)(G_{11}\psi_1 c_2 - G_{12}\psi_2 c_1) \quad (32)$$

$$B^0 = c_1(c_1\tau_2 + c_2\tau_1 - \frac{1}{2}) + \psi_2'(c_1\tau_2 + c_2\tau_1)(G_{22}\psi_2 c_1 - G_{12}\psi_1 c_2). \quad (33)$$

In a pure diffusion process, if we assume the sound speed is much larger than the speed of diffusion, the pressure field can be taken as uniform. When the equation of state is not an ideal gas one, the total density field is not a uniform constant during the diffusion. The density perturbations $\rho_\sigma^1$ are therefore not independent of each other, and a velocity field with a generally non-zero divergence will be generated. In the present case, $\rho_1^1$ and $\rho_2^1$ are related by Eq. (19):

$$[1 + \psi_1'(G_{11}\psi_1 + G_{21}\psi_2)]\rho_1^1 + [1 + \psi_2'(G_{12}\psi_1 + G_{22}\psi_2)]\rho_2^1 = 0. \quad (34)$$

We can then write $\rho_1^1$ and $\rho_2^1$ in terms of a single perturbation field, $\xi(\mathbf{x})$,

$$\rho_1^1(\mathbf{x}) = [1 + \psi_2'(G_{12}\psi_1 + G_{22}\psi_2)]\xi(\mathbf{x}) \quad (35)$$

$$\rho_2^1(\mathbf{x}) = -[1 + \psi_1'(G_{11}\psi_1 + G_{21}\psi_2)]\,\xi(\mathbf{x}). \tag{36}$$

It follows from the continuity equation (25) that

$$[\psi_2'(G_{12}\psi_1 + G_{22}\psi_2) - \psi_1'(G_{11}\psi_1 + G_{21}\psi_2)]\frac{D\xi}{Dt} + \rho^0 \nabla \cdot \mathbf{u} = 0. \tag{37}$$

On substituting the three equations above into either Eq. (30) or (31), we find that $\xi(\mathbf{x})$ satisfies the following diffusion equation

$$\frac{D\xi}{Dt} = \frac{c^2(1-d_0)}{D}\left[(c_1\tau_2 + c_2\tau_1)\gamma - \frac{1}{2}\right]\nabla^2\xi, \tag{38}$$

where, the effect of long range interaction on the diffusivity is absorbed into the single factor

$$\gamma = \frac{(1 + G_{11}\psi_1\psi_1')(1 + G_{22}\psi_2\psi_2') - G_{12}^2\psi_1\psi_1'\psi_2\psi_2'}{1 + c_1 G_{11}\psi_1\psi_1' + c_2 G_{22}\psi_2\psi_2' + G_{12}\left(c_1\psi_1'\psi_2 + c_2\psi_2'\psi_1\right)}. \tag{39}$$

$\gamma$ becomes unity in the absence of long range interaction. Furthermore, if the relaxation times of the two components are equal, namely $\tau_1 = \tau_2 = \tau$, the diffusivity becomes $c^2(1-d_0)(\tau - 1/2)/D$, in agreement with other authors.[13,20]

A few remarks are called for at this point upon the diffusivity derived above. First, the diffusivity depends on the relative density concentration of the two components as expected. This dependence, through not only the $c_\sigma$'s but also the $\psi_\sigma$'s, is quite general and can theoretically be tuned to approximate any given diffusivity. We are therefore able to simulate a class of diffusion problems in which the diffusivity can vary and depend on concentrations. Second, unlike the LBE model for miscible fluid due to Flekkøy,[20] in which the mutual diffusivity depends on the velocity of the fluid, the diffusivity does not have a velocity dependence. The diffusion in the present LBE model is fully Galilean invariant.

When applied to convection-diffusion problems, we notice that the diffusivity can be varied independently of the shear viscosity by changing the long range interaction potential. Generally attractive forces between particles of same component, and repulsive forces between different components both tend to decrease the diffusivity.

The second order momentum equation can be derived in a way similar to that of one-component fluid without interparticle interaction.[19] It turns out that long range interaction does not affect the shear viscosity, at least to second order. When multiple components are present, the shear viscosity is simply $\nu = c^2(\sum_\sigma c_\sigma \tau_\sigma - \frac{1}{2})/(D+2)$.

## 3. Numerical simulation

In this section, we present results of numerical simulations of the LBE model described above. We first measure the mutual diffusivity in a binary mixture by studying the decay of a sinusoidal concentration wave with small amplitude. The simulation was carried out on a 256 × 16 hexagonal lattice. Initially the perturbations of the densities of the two components were set up according to Eq. (34) so that the pressure was uniform in the whole field. The relaxation times of the two components are $\tau_1 = 0.7$ and $\tau_2 = 1.0$. Both number densities were set equal to unity. The "effective masses" were chosen to be $\psi_1(n_1) = 1 - \exp(-n_1)$ and $\psi_2 = n_2$. The diffusivity is obtained by measuring the decay rate of the concentration wave. The results of the measurement are plotted in Fig. 1 as function of $G_{12}$, which measures the strength of the force between particles of the two components. $G_{11}$ and $G_{22}$ are fixed at positive numbers 0.01 and 0.02 respectively, so that interactions between all the particles are repulsive. Two sets of data are presented for two different values of the molecular mass of the second component, $m_2 = 2$ and $m_2 = 4$. The first component always has a molecular mass of 1.0. The density concentration $c_1$ is then 0.33 and 0.2 in the two cases. In Fig. 1, the solid lines are the analytic predictions, which are in excellent agreement with the experimental values.

The mutual diffusivity is also measured in the presence of a uniform velocity field parallel to the concentration gradient. No dependence of the diffusivity on the velocity is observed. For instance, when all the parameters are chosen to be the same as in the case above in which $m_2 = 2$ and $G_{12} = 0.04$, the diffusivity measured without flow is 0.0549, compared with 0.0550 when a uniform flow of speed 0.2 is present.

The expression for the shear viscosity is also verified similarly by measuring the decay rate of a sinusoidal transverse wave on the same lattice. The concentrations of the two components are varied by changing the molecular mass ratio while keeping the total density of the fluid constant. Without loss of generality, we chose the number densities $n_1 = n_2 = 1$ and $\rho = 5$ in the simulation. The relaxation times of the two components are 0.6 and 1.2 respectively. $G_{12}$ equals 0.04 here and all other parameters are the same as before. The measured viscosity is plotted in Fig. 2 as function of the density concentration of the first component. It is again in excellent agreement with the analytic prediction indicated by the solid straight line.

## 4. Conclusions

We have described in detail a multiple-component LBE model with interparticle interactions. The equations governing the evolution of the densities of the components are derived using the Chapman-Enskog technique. The mutual diffusivity and shear viscosity in a binary mixture are derived analytically using those equations and the agreement is obtained to high accuracy with numerical simulations. The diffusivity generally has an explicit dependence on the concentrations of the two components and can be tuned to model physical diffusivities which are concentration-dependent. When the interparticle interaction is turned off and the relaxation times of the two components are equal, the diffusivity does not depend on the concentrations, in accordance with other authors. The Galilean invariance of the diffusivity in this model is also confirmed by numerical simulations.

In the current paper, we focus our attention to the derivation of macroscopic equations and the transport coefficients of multiple-component system, which can be fully or partially miscible. In the latter case, this model can be used to simulate fluid flows which involve phase transition and interfaces between different phases. The derivation of the transport coefficients in this case becomes more valuable because in some cases the concentration gradients near an interface are so large that the diffusivity can not be approximated as a constant. The dependence of the diffusivity on the concentration itself is essential in the study of convection-diffusion problems that involve interface. In principal, the phase-diagram, the density-profile across an interface, and the surface tension can be calculated for a multiple-component system in a way similar to, but more complex than, a one-component system.[17] We defer the detailed discussion to a future publication.

**Acknowledgements**

The authors thank Dr. Hudong Chen and Dr. David Montgomery for helpful discussion. This work is supported in part at Dartmouth by subcontract No. 9-XA3-1416E from Los Alamos National Laboratory,

**Figures**

Fig. 1. Mutual diffusivity in a binary mixture, as function of the strength of the repulsive force between the two components. Solid lines are theoretical predictions and the symbols are the results of numerical simulation. The two sets data are for two different density ratios of the two components.

Fig. 2. Shear viscosity as function of the density of the first component. Interparticle interactions exist in the simulation but this has no effect on the shear viscosity. The solid line is the theoretical prediction.

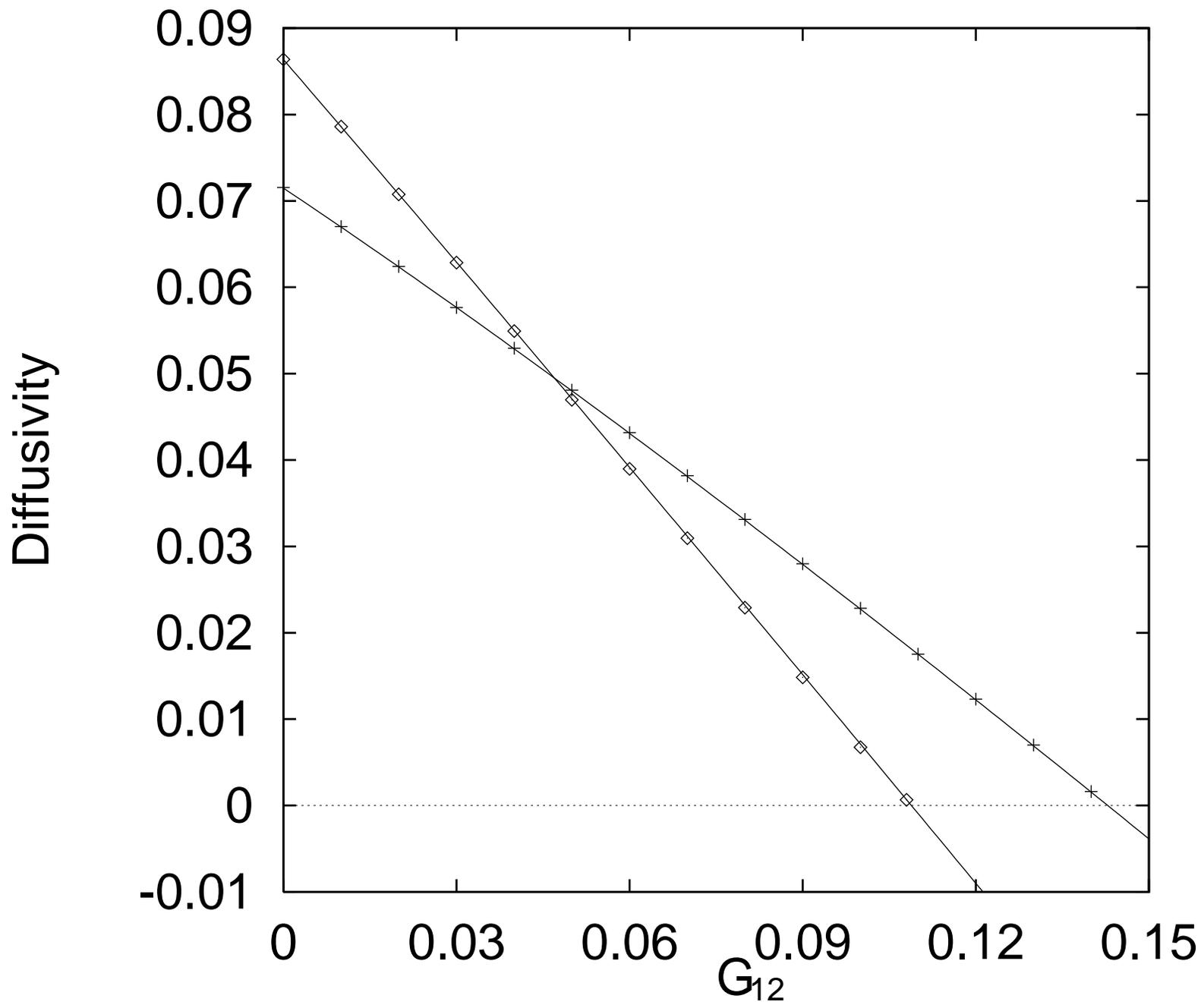

Figure 1.

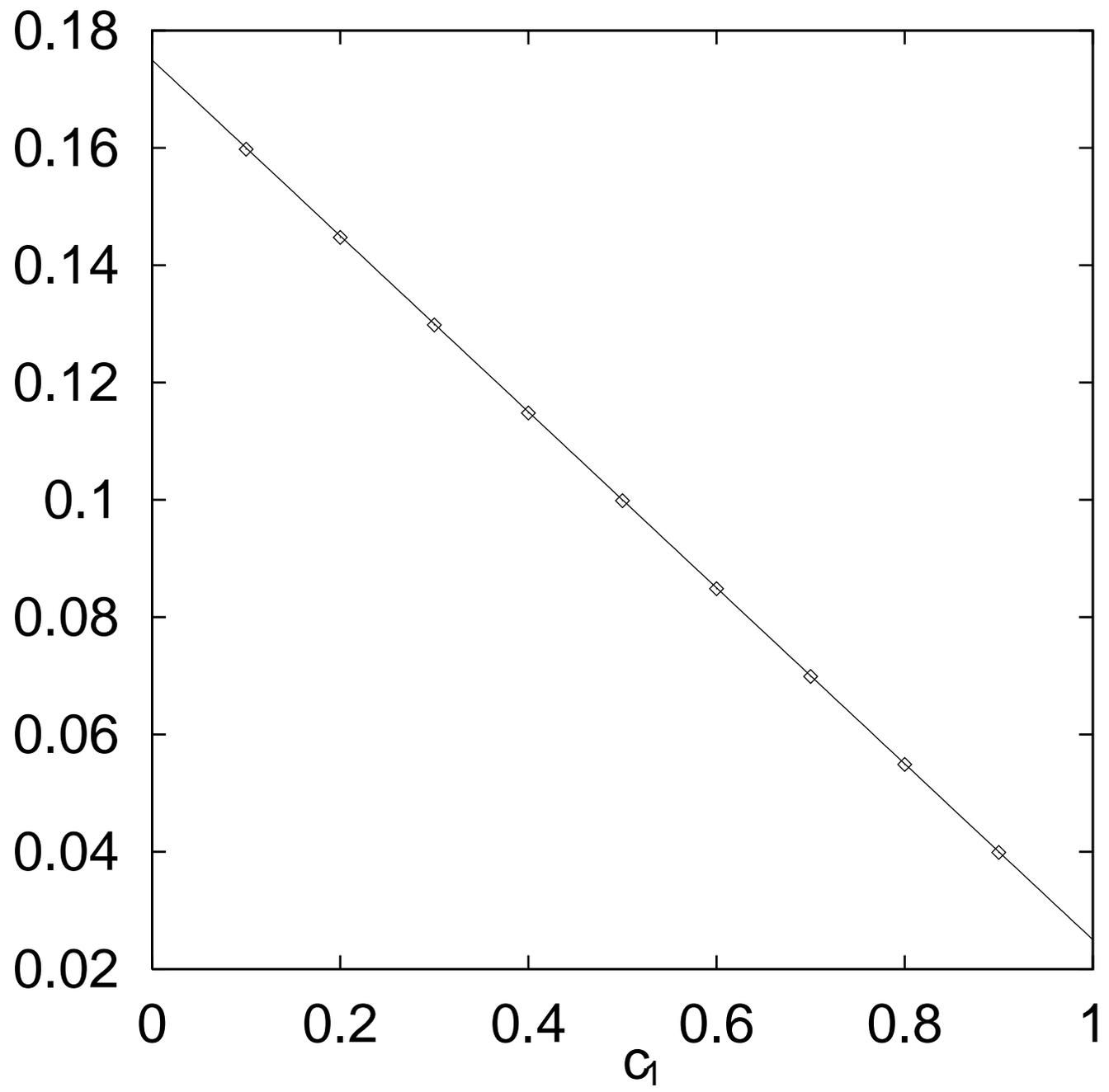